\documentclass{ieeeaccess}
\usepackage{cite}
\usepackage{amsmath,amssymb,amsfonts,amsthm}
\usepackage{graphicx}
\usepackage{textcomp}
\usepackage{booktabs}
\usepackage{algorithm}
\usepackage{algpseudocode}
\usepackage{xcolor}
\usepackage[hidelinks,bookmarks=false]{hyperref}
\newtheorem{proposition}{Proposition}
\newtheorem{corollary}{Corollary}
\theoremstyle{remark}
\newtheorem{remark}{Remark}
\providecommand\xfigwd{0pt}
\makeatletter 
\ifdefined\c@biography\else\fi
\ifdefined\if@biographyTOCentrynotmade\else\newif\if@biographyTOCentrynotmade\@biographyTOCentrynotmadetrue\fi
\makeatother
\newcommand{\dayGainPct}{7.5}
\newcommand{\dayOracleFrac}{74}
\newcommand{\nightAdaptFrac}{99.8}
\newcommand{\dawnAdaptFrac}{100.1}
\newcommand{\dayFourLoss}{38}
\newcommand{\nightEightLoss}{26}
\newcommand{\daySwitches}{12.7}

\newcommand{\NSEED}{50}

\newcommand{\nBlocks}{160}
\newcommand{\hwSlopeMin}{0.89}
\newcommand{\hwSlopeMax}{1.18}
\newcommand{\hwSpread}{1.5}
\newcommand{\offTwoZ}{0.7}
\newcommand{\offTwoX}{1.1}
\newcommand{\offFourZ}{2.8}
\newcommand{\offFourX}{2.5}
\newcommand{\offEightZ}{6.7}
\newcommand{\offEightX}{7.6}

\newcommand{\hwJob}{dbvplp5s2omc73889n0g}
\newcommand{\twoqTwo}{1}
\newcommand{\twoqFour}{5}
\newcommand{\twoqEight}{19}
\newcommand{\loopDayDrop}{59}

\newcommand{\kfGainMin}{0.5}
\newcommand{\kfGainMax}{1.0}

\newcommand{\raGainMin}{31}
\newcommand{\raGainMax}{47}
\newcommand{\recalWorst}{98.8}

\newcommand{\fixSixteenWorst}{63.8}
\newcommand{\fixBestWorst}{73.5}

\newcommand{\recalMMgainMax}{13}
\newcommand{\recalCostMin}{0.4}
\newcommand{\recalCostMax}{4.7}
\newcommand{\epsPass}{1.6\times10^{-8}}

\newcommand{\Hd}{H_d}

\newcommand{\epsec}{\varepsilon_{\rm sec}}
\newcommand{\epcor}{\varepsilon_{\rm cor}}

\begin{document}
\history{This work has been submitted to the IEEE for possible publication. Copyright may be transferred without notice, after which this version may no longer be accessible.}
\doi{}

\title{Feed-Forward Dimension Control for High-Dimensional Quantum Key Distribution over Time-Varying Airborne Free-Space Links}

\author{\uppercase{Murtuza Lemonwala}\authorrefmark{1}
and \uppercase{Laxima Niure Kandel}\authorrefmark{1}}
\address[1]{Department of Electrical Engineering and Computer Science, Embry-Riddle Aeronautical University, Daytona Beach, FL 32114 USA}

\tfootnote{This work did not receive any funding. The authors acknowledge the use of IBM Quantum services for this work.}

\markboth
{Lemonwala and Niure Kandel: Feed-Forward Dimension Control for High-Dimensional QKD}
{Lemonwala and Niure Kandel: Feed-Forward Dimension Control for High-Dimensional QKD}

\corresp{Corresponding author: Murtuza Lemonwala (email: lemonwam@my.erau.edu).}

\begin{abstract}
High-dimensional quantum key distribution tolerates more noise than qubit-based protocols, but the dimension that maximizes the secret key depends on turbulence, background light, device imperfections, and block size, all of which change during an airborne or satellite pass. Finite-size security analyses of high-dimensional protocols fix the dimension in advance, and the dimension-adaptive schemes reported so far have been analyzed only asymptotically. Here, a controller sets the dimension, acceptance thresholds, and key length of each block using only public data from earlier blocks. Each block is therefore a standard fixed-length protocol, and composable finite-key security against coherent attacks follows from sequential composition; an inaccurate channel model reduces the key rate but does not weaken security. The controller estimates turbulence from observed error rates, maximizes the expected key under forecast uncertainty, and recalibrates its model online. We evaluate it on a simulated entanglement-based orbital-angular-momentum link between an aircraft and a ground station, using Kolmogorov phase screens and the Hufnagel--Valley turbulence profile. Accounting for forecast uncertainty increases the key by 31\%--47\% relative to a naive planner. Over night, dawn, daytime, and model-mismatch scenarios, the recalibrated controller retains at least 98.8\% of the key obtained with the best fixed dimension for each scenario, while every fixed dimension drops to 73.5\% or less in at least one scenario. Running the link's measurement statistics on a superconducting quantum processor shows device errors close to a constant offset; with these errors included, the best daytime dimension changes from eight to four, and the controller adjusts accordingly.
\end{abstract}


\titlepgskip=-15pt

\maketitle

\section{Introduction}\label{sec:intro}
\PARstart{Q}{uantum} key distribution (QKD) allows two parties to establish a shared secret key with security based on the laws of quantum mechanics rather than on computational hardness~\cite{BB84,BBM92,Scarani2009,Pirandola2020}. Free-space QKD extends this to mobile platforms and to locations without fiber. Air-to-ground links have been demonstrated from aircraft~\cite{Nauerth2013,Pugh2017} and drones~\cite{Liu2021drone}, and satellites have distributed keys and entanglement over distances above 1000\,km~\cite{Liao2017,Yin2017}. In all of these links, the channel changes during a pass: the range, elevation angle, turbulence strength, and background light vary on time scales of seconds.

Encoding each photon in $d>2$ dimensions carries up to $\log_2 d$ bits per detected photon and increases the error rate that the protocol can tolerate~\cite{Cerf2002,Sheridan2010,Cozzolino2019,Erhard2020}. With two mutually unbiased bases (MUBs) and symmetric noise, the asymptotic error threshold is 11.0\% at $d=2$ and rises to 18.9\%, 24.7\%, and 29.0\% at $d=4$, 8, and 16 (Table~\ref{tab:thresh}). For this reason, high-dimensional (HD) QKD has been studied for free-space links, where background light and turbulence are the main impairments~\cite{Mafu2013,Mirhosseini2015,Krenn2014,Bulla2023,Scarfe2025}.

Larger $d$ is not always better. With spatial-mode encodings, turbulence-induced crosstalk increases with mode order~\cite{Paterson2005}, and the background count rate increases with the number of monitored modes. Device imperfections and finite-size effects at short block lengths also favor smaller $d$. The optimal dimension thus depends on the state of the link, which on a moving platform changes within seconds. Dimension-adaptive operation has been suggested for such non-static links~\cite{Ecker2023space,Ecker2021}, and a dimensionally adaptive protocol was analyzed for a 10.2\,km free-space experiment, although only in terms of asymptotic key rates~\cite{Bulla2023}.

Composable finite-size security proofs for general HD-QKD protocols have appeared only recently~\cite{KanitscharPRApplied2025,KanitscharPRL2025} and were applied at a fixed dimension of $d=16$. In a large time--frequency experiment, Chang \emph{et al.}~\cite{Chang2026} found that the finite-size-optimal dimension depends strongly on block size and on the attack model; the dimension was selected by comparing key rates after the measurement, and the authors pointed out that in continuous operation it would normally have to be set beforehand. Kanitschar~\cite{KanitscharThesis2025} likewise lists the a priori choice of the system dimension as an open problem. Current free-space HD experiments set the dimension before deployment~\cite{Dziwulska2026}, and recent noise models~\cite{BergmayrMann2026} and optimization schemes for spatial-mode encodings~\cite{Bhat2025} select it in advance using asymptotic rates.

In this article we study how to change the dimension during operation while keeping composable finite-key security. The approach rests on the following observation. If the parameters of each block, including the dimension, are fixed using only public data from earlier blocks, each block is a standard fixed-length protocol and its security follows from existing proofs through sequential composition. Channel modeling, forecasting, and the handling of risk then influence the amount of key produced but play no role in the security argument. We design a controller based on this principle and test it on a detailed model of an aircraft-to-ground link and on a superconducting quantum processor.

\subsection{Contributions}
\begin{enumerate}
\item \emph{Protocol and security.} We define feed-forward dimension control (Section~\ref{sec:ff}), in which the dimension, acceptance thresholds, and key length of block $j$ are computed from the public transcript of blocks $1,\dots,j-1$. Security against coherent attacks then follows from fixed-length proofs and sequential composition (Proposition~\ref{prop:ff}). We write out the hybrid argument in full because it applies equally to other data-driven parameter controllers, including those based on machine learning~\cite{WangLo2019,OptiQKD2026}.
\item \emph{Controller.} The controller estimates the turbulence strength from observed error rates through a channel model, tracks its variability, chooses the dimension and thresholds that maximize the expected key, and corrects its model using the error rates it observes. An ablation study quantifies the contribution of each part.
\item \emph{Airborne link study.} Using Kolmogorov phase screens and a simulated aircraft-to-ground pass, we compare the controller with fixed dimensions, measure the loss due to causal forecasting against a non-causal oracle and a Kalman-filter forecaster, and examine the effect of block duration and of errors in the channel model. We also map the finite-key-optimal dimension over turbulence and background levels.
\item \emph{Processor emulation.} We run the measurement statistics of the link on a superconducting quantum processor and include the measured device errors in the link simulation. These errors change which dimension is optimal.
\end{enumerate}

\subsection{Organization}
Section~\ref{sec:related} reviews related work. Section~\ref{sec:system} defines the link, protocol, and assumptions. Section~\ref{sec:keyrate} gives the per-block finite-key length. Section~\ref{sec:ff} presents feed-forward control, its security, and the controller. Section~\ref{sec:model} describes the turbulence and pass model, and Section~\ref{sec:hw} the processor emulation. Section~\ref{sec:results} reports results, Section~\ref{sec:discussion} discusses implementation and limitations, and Section~\ref{sec:conclusion} concludes. Appendix~\ref{app:repro} summarizes reproducibility details.

\section{Related Work}\label{sec:related}
\subsection{High-Dimensional QKD and Finite-Key Security}
Cerf \emph{et al.}~\cite{Cerf2002} showed that qudit protocols tolerate more noise than qubit protocols, and Sheridan and Scarani~\cite{Sheridan2010} proved their security and found that finite-key corrections depend only weakly on $d$. The entropic uncertainty relation for smooth entropies~\cite{TomamichelRenner2011} leads to tight finite-key bounds against coherent attacks~\cite{Tomamichel2012}. These bounds were extended to multilevel protocols with two MUBs~\cite{Bradler2016} and used in a time-bin qudit implementation~\cite{Islam2017}. Kanitschar and Huber~\cite{KanitscharPRL2025,KanitscharPRApplied2025} developed a composable finite-size framework for general HD-QKD protocols, including protocols that use fewer than a full set of MUBs. In all of these analyses, $d$ is a fixed parameter of the protocol.

\subsection{Free-Space and Airborne Links}
Orbital angular momentum (OAM) modes~\cite{Allen1992} provide a large alphabet for free-space encoding and can be sorted with high efficiency~\cite{Berkhout2010}. OAM-based HD-QKD with MUBs has been demonstrated in the laboratory~\cite{Mafu2013,Mirhosseini2015}, and spatial modes have been sent through several kilometers of turbulent air~\cite{Krenn2014}. Because turbulence couples neighboring OAM orders, crosstalk increases with $d$~\cite{Paterson2005}. Adaptive optics reduces this crosstalk~\cite{Scarfe2025}, and neural-network predictors of turbulence for free-space QKD have been proposed~\cite{Jaouni2024}. Time-bin encodings are less sensitive to turbulence and offer an alternative~\cite{Bulla2023,Islam2017}. For satellite links, finite-block effects over a pass have been quantified~\cite{Sidhu2022}, day and night data have been processed in separate blocks~\cite{Bae2023}, and strategies for increasing key rates have been reviewed~\cite{Ecker2021}.

\subsection{Adaptive and Data-Driven QKD}
Machine-learning and reinforcement-learning methods have been used to tune decoy intensities and other parameters in real time~\cite{WangLo2019,OptiQKD2026}. Variable-length protocols allow the key length to depend on the data of the same block~\cite{Tupkary2024}, and marginal-constrained entropy accumulation allows the statistical analysis of later rounds to depend on earlier public announcements~\cite{ArqandTan2025,TupkaryThesis2026}. None of these works changes the measurement dimension during operation, and the learning-based methods do not address how their data-dependent choices affect composable security. The present work combines dimension switching, a composable finite-key argument, and a controller evaluated on a detailed physical model of the link.

\section{System Model}\label{sec:system}
\subsection{Link Architecture}
\begin{figure*}[t]
\centering
\includegraphics[width=\textwidth]{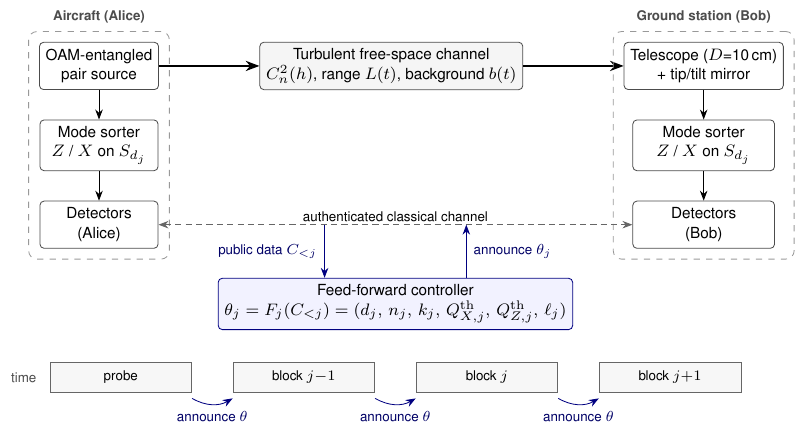}
\caption{System overview. An airborne source distributes OAM-entangled pairs to Alice (on board) and Bob (ground). Before block $j$, the controller fixes all block parameters $\theta_j$ from the public transcript $C_{<j}$ of earlier blocks and announces them; block $j$ is then run as a standard fixed-length protocol in dimension $d_j$. Block 0 is a probe that produces no key.}\label{fig:system}
\end{figure*}

We consider an entanglement-based link (Fig.~\ref{fig:system}). A source on board the aircraft emits OAM-entangled photon pairs. One photon is measured on board (Alice), and its partner is transmitted to a ground receiver (Bob). Bob's telescope is followed by a fast-steering mirror that removes tip and tilt. Both parties use mode sorters~\cite{Berkhout2010} that project onto Laguerre--Gauss modes $\mathrm{LG}_{0}^{\ell}$ with azimuthal index $\ell$. An authenticated classical channel carries the public discussion.

For a dimension $d$ in a finite menu $\mathcal D=\{2,4,8,16\}$, the active mode set is
\begin{equation}
S_d=\{\ell\in\mathbb Z:-d/2\le\ell<d/2\}.
\end{equation}
These sets are nested, $S_2\subset S_4\subset S_8\subset S_{16}$, so a receiver that resolves all 16 modes can switch dimension by changing which outcomes it keeps and which superpositions it measures. The key basis $Z$ is the OAM basis $\{|\ell\rangle\}_{\ell\in S_d}$. The test basis $X$ is its discrete Fourier (``ANGLE'') basis,
\begin{equation}
|\tilde m\rangle=\frac{1}{\sqrt d}\sum_{\ell\in S_d}\omega^{m\,\mathrm{idx}(\ell)}|\ell\rangle,\qquad \omega=e^{2\pi i/d},
\end{equation}
where $\mathrm{idx}(\ell)\in\{0,\dots,d-1\}$ indexes $S_d$. The two bases are mutually unbiased.

\subsection{Protocol}
Time is divided into blocks of duration $T$. Block $j$ proceeds as follows.
\begin{enumerate}
\item \emph{Announcement.} The block parameters
\[\theta_j=(d_j,n_j,k_j,Q_{X,j}^{\rm th},Q_{Z,j}^{\rm th},\ell_j)\]
are fixed from the public transcript $C_{<j}$ of earlier blocks and announced over the authenticated channel.
\item \emph{Measurement.} Each party independently measures each photon in $Z$ with probability $p_Z$ and in $X$ otherwise, keeping only detections in $S_{d_j}$. Detections outside $S_{d_j}$ are discarded as loss.
\item \emph{Sifting.} The parties announce bases. Rounds in which both chose $Z$ form the raw key, and rounds in which both chose $X$ form the test set.
\item \emph{Parameter estimation.} The parties compare the test outcomes publicly. The block aborts if fewer than $n_j$ key rounds or $k_j$ test rounds were collected, or if the observed test error exceeds $Q_{X,j}^{\rm th}$.
\item \emph{Error correction and verification.} Bob corrects his raw key with a code designed for error rate $Q_{Z,j}^{\rm th}$; a hash comparison verifies the result. The block aborts if verification fails, which we model as the observed $Z$ error exceeding $Q_{Z,j}^{\rm th}$.
\item \emph{Privacy amplification.} Two-universal hashing produces a key of the pre-announced length $\ell_j$.
\end{enumerate}
The public transcript $C_j$ of block $j$ comprises $\theta_j$, the basis announcements, the test outcomes, the error-correction syndrome and hash, the accept/abort flag, and the total detection count. Because $X$ and $Z$ span the same $d$-dimensional subspace, the discard event corresponds to the same positive-operator-valued-measure element $\mathbb 1-P_{S_d}$ in both bases. The post-selection is therefore independent of the basis choice, as required by the uncertainty-relation proofs used in Section~\ref{sec:keyrate}.

\subsection{Threat Model and Assumptions}
Eve has full control of the quantum channel and may perform arbitrary coherent attacks, including attacks that correlate blocks and adapt to every public announcement. She reads, but cannot alter, the authenticated classical channel. As is standard for device-dependent proofs~\cite{Scarani2009,Tomamichel2012}, we assume that Alice's and Bob's measurements are the specified MUB projections on $S_d$, that detection is basis-independent, and that the devices are trusted and characterized. Side channels, detector dead time, and multipair emission are not considered (Section~\ref{sec:limits}).

\section{Per-Block Finite-Key Length}\label{sec:keyrate}
Consider a single fixed-length block with $n$ key rounds, $k$ test rounds, test-error threshold $Q_X^{\rm th}$, and error-correction code designed for $Q_Z^{\rm th}$. Following the uncertainty-relation approach of~\cite{TomamichelRenner2011,Tomamichel2012} in the qudit two-MUB form of~\cite{Bradler2016}, the block yields an $\epsec$-secret and $\epcor$-correct key of length
\begin{multline}
\ell=\Big\lfloor n\big[\log_2 d-\Hd(Q_X^{\rm th}+\mu)\big]-f\,n\,\Hd(Q_Z^{\rm th})\\
-\log_2\frac{2}{\epsec^2\,\epcor}\Big\rfloor .\label{eq:ell}
\end{multline}
Here
\begin{align}
\Hd(x)&=-x\log_2\frac{x}{d-1}-(1-x)\log_2(1-x)\nonumber\\
&=h(x)+x\log_2(d-1)
\end{align}
bounds the per-symbol log-volume of a $d$-ary Hamming ball, with $h$ the binary entropy; $\log_2 d$ is the uncertainty-relation constant for two MUBs; $f\ge1$ is the error-correction inefficiency; and
\begin{equation}
\mu=\sqrt{\frac{n+k}{nk}\,\frac{k+1}{k}\,\ln\frac{2}{\epsec}}\label{eq:mu}
\end{equation}
is the random-sampling correction that bounds the phase error rate on the key rounds given the observed test error~\cite{Tomamichel2012,Bradler2016}. The bound is stated for $Q_X^{\rm th}+\mu\le 1/2$~\cite{Bradler2016}; our implementation enforces this, and returns zero key otherwise.

With all parameters $\theta=(d,n,k,Q_X^{\rm th},Q_Z^{\rm th},\ell)$ fixed in advance, such a block is $(\epsec+\epcor)$-secure against coherent attacks, because the uncertainty relation holds for arbitrary quantum side information~\cite{TomamichelRenner2011,Tomamichel2012}. Sheridan and Scarani found finite-key corrections of this kind to be almost independent of $d$ for dimensions up to about 20~\cite{Sheridan2010}, which covers the menu used here. To verify the implementation, we confirmed that the asymptotic form of \eqref{eq:ell}, $r=\log_2 d-\Hd(Q_X)-\Hd(Q_Z)$ with $Q_X=Q_Z$, gives the known two-MUB thresholds (Table~\ref{tab:thresh}).

\begin{table}[t]
\caption{Asymptotic error thresholds of the two-MUB protocol from \eqref{eq:ell} with $f=1$, $Q_X=Q_Z$}\label{tab:thresh}
\centering\footnotesize
\begin{tabular}{@{}lcccc@{}}
\toprule
Dimension $d$ & 2 & 4 & 8 & 16 \\
\midrule
Bits per key symbol, $\log_2 d$ & 1 & 2 & 3 & 4 \\
Error threshold $e^\ast$ (\%) & 11.00 & 18.93 & 24.70 & 28.97 \\
\bottomrule
\end{tabular}
\end{table}

Throughout we use the basis bias $p_Z=0.9$, so that $n=Np_Z^2$ and $k=N(1-p_Z)^2$ for $N$ kept coincidences; the security parameters $\epsec=10^{-10}$ and $\epcor=10^{-15}$; and $f=1.16$, typical of nonbinary low-density parity-check reconciliation for HD-QKD~\cite{Muller2023}.

\section{Feed-Forward Dimension Control}\label{sec:ff}
\subsection{Security of Feed-Forward Selection}\label{sec:security}
The following result follows directly from the composability of QKD security~\cite{Renner2005,PortmannRenner2022}. We include a proof because the result is what allows the measurement dimension, or any other parameter chosen from data, to change between blocks without a new security analysis.

\begin{proposition}\label{prop:ff}
Let $\Theta$ be a finite parameter set such that each fixed-length block protocol $\Pi_\theta$, $\theta\in\Theta$, is $\varepsilon_\theta$-secure against arbitrary attacks, with $\varepsilon_\theta\le\varepsilon$. Suppose block $j$ runs $\Pi_{\theta_j}$ with $\theta_j=F_j(C_{<j})$, where $F_j$ is any deterministic function of the public transcript $C_{<j}$ of earlier blocks. Then the $J$ keys produced are jointly $J\varepsilon$-secure.
\end{proposition}

\begin{proof}
Let $\rho^{\rm real}$ denote the joint state of all keys $K_1,\dots,K_J$, the transcript $C=(C_1,\dots,C_J)$, and Eve's quantum system $E$ at the end of the protocol. Let $\rho^{(i)}$ be the hybrid in which the keys of blocks $1,\dots,i$ are replaced by their ideal versions (uniformly random and identical for Alice and Bob whenever the block does not abort, and independent of everything else), with $\rho^{(0)}=\rho^{\rm real}$ and $\rho^{(J)}=\rho^{\rm ideal}$.

Compare $\rho^{(i-1)}$ with $\rho^{(i)}$. Condition on any value $c_{<i}$ of the classical transcript before block $i$. Then $\theta_i=F_i(c_{<i})$ is a fixed element of $\Theta$, and block $i$ executes $\Pi_{\theta_i}$ on whatever joint state Eve has prepared, which may be correlated arbitrarily with earlier blocks, earlier keys, and her own memory. The security definition of $\Pi_{\theta}$ quantifies over all such input states~\cite{PortmannRenner2022}, so the real and ideal outputs of block $i$ are $\varepsilon$-close in trace distance for every $c_{<i}$. Because $C_{<i}$ is a classical register that appears identically in both hybrids, the trace distance of the conditional mixture is the average of the conditional distances and is therefore at most $\varepsilon$. Everything that happens after block $i$, including the choice of $\theta_{i+1},\dots,\theta_J$ and the execution of later blocks, is a single completely positive trace-preserving map applied identically to both hybrids, because it depends only on public data and on systems that the two hybrids share; it cannot increase the trace distance. Hence $\tfrac12\|\rho^{(i-1)}-\rho^{(i)}\|_1\le\varepsilon$, and the triangle inequality over $i=1,\dots,J$ gives $\tfrac12\|\rho^{\rm real}-\rho^{\rm ideal}\|_1\le J\varepsilon$.
\end{proof}

\begin{corollary}\label{cor:pass}
With $\epsec=10^{-10}$, $\epcor=10^{-15}$, and $J=160$ blocks per pass, each pass is $\epsPass$-secure, independently of the channel model, the forecaster, and the risk level used by the controller.
\end{corollary}

\begin{remark}[Effect of the controller]
The channel model, the forecast, and the risk level determine which protocol is run in each block, but every candidate protocol is secure. An inaccurate model therefore lowers the key rate without affecting security. Since $C_{<j}$ can be influenced by Eve, she can push the controller toward poor choices; this amounts to a denial-of-service attack, which she can already mount by blocking the channel.
\end{remark}

\begin{remark}[Scope]
If the dimension or key length were chosen after seeing the data of the same block, as when key rates are compared in post-processing, fixed-length proofs would no longer apply. A union bound over the menu of dimensions or a variable-length argument~\cite{Tupkary2024} would then be needed. Entropy-accumulation methods allow the statistical analysis to adapt round by round~\cite{ArqandTan2025,TupkaryThesis2026}; combining this with dimension switching remains an open problem.
\end{remark}

\subsection{Controller}\label{sec:controller}
The controller (Algorithm~\ref{alg:ff}) runs between blocks. Its inputs are the public data of block $j-1$: the detection count $N$, the error rates $\hat e_Z$ and $\hat e_X$ obtained in the public comparison, and the background estimate $b$. Its components are described below, and its settings are listed in Table~\ref{tab:hyper}.

\subsubsection{Turbulence inversion}
The channel model provides ensemble-averaged error rates $e_B^{\rm ch}(d,x)$ for $B\in\{Z,X\}$ and in-set transmission $\eta(d,x)$ as functions of the turbulence strength $x=D/r_0$ (Section~\ref{sec:model}). With $N_b=d\,b\,T$ expected background counts, uniformly spread over the $d$ outcomes, the signal count and the signal-only error rate at the current dimension are estimated as
\begin{equation}
\hat S=\max(N-N_b,1),\quad \hat e^{\rm ch}_Z=\frac{\hat e_Z N-N_b(d-1)/d}{\hat S}.\label{eq:inv1}
\end{equation}
The turbulence estimate is the model inverse,
\begin{equation}
\hat x=\arg\min_{x\in[0,6]}\big|e_Z^{\rm ch}(d,x)-\hat e^{\rm ch}_Z\big|,\label{eq:inv2}
\end{equation}
evaluated on a grid of 2401 points.

\subsubsection{Forecasting}
For a candidate dimension $d'$ and turbulence level $x'$, the controller forecasts
\begin{align}
S_{d'}&=\hat S\,\frac{\eta(d',x')}{\eta(d,\hat x)},\qquad N_{d'}=S_{d'}+d'bT,\\
e_{B,d'}&=\frac{S_{d'}\,e_B^{\rm ch}(d',x')+d'bT\,(d'-1)/d'}{N_{d'}}.\label{eq:fc}
\end{align}

\subsubsection{Fixed-length planning}
Given a forecast $(N,e_Z,e_X)$, the planner derates the count to $0.97N$, sets $n$ and $k$ accordingly, and sets the acceptance thresholds with a $3$-standard-deviation sampling margin and a small absolute margin,
\begin{equation}
Q_X^{\rm th}=e_X+3\sqrt{\frac{e_X(1-e_X)}{k}}+0.002,\label{eq:th}
\end{equation}
and likewise $Q_Z^{\rm th}$ with $n$ in place of $k$. The key length $\ell$ then follows from \eqref{eq:ell}. A candidate is discarded if $n$ or $k$ is below 10, if $\ell\le0$, or if the validity condition of Section~\ref{sec:keyrate} fails.

\subsubsection{Risk-aware selection}
Because the turbulence changes from one block to the next, thresholds that include only a statistical margin would cause many blocks to abort. The controller therefore estimates the block-to-block variability $\sigma$ of $\ln\hat x$ with an exponentially weighted moving average (EWMA),
\begin{equation}
\sigma\leftarrow0.8\,\sigma+0.2\cdot1.25\,\big|\ln\hat x_{j-1}-\ln\hat x_{j-2}\big|,\label{eq:sigma}
\end{equation}
where the factor $1.25\approx\sqrt{\pi/2}$ converts a mean absolute deviation into a standard deviation for Gaussian increments. For each pessimism level $z\in\{0,0.5,\dots,3\}$, it plans at the pessimistic turbulence level $x'=\hat x\,e^{z\sigma}$ and scores the plan by its expected key,
\begin{equation}
V(d',z)=\ell(d',z)\,\Phi(z),\label{eq:V}
\end{equation}
where $\Phi$ is the standard normal cumulative distribution function, an estimate of the probability that the block does not abort. The plan with the largest $V$ over $(d',z)$ is announced.

\subsubsection{Online recalibration}
The turbulence estimate is only as accurate as the channel model used to obtain it. To correct for model error, the controller keeps per-dimension correction factors and updates them after each block run at dimension $d$:
\begin{equation}
\rho_d^{B}\leftarrow0.7\,\rho_d^{B}+0.3\,\mathrm{clip}\!\left(\frac{\hat e_B}{e_{B}^{\rm pred}},\,0.5,\,3\right),\quad B\in\{Z,X\},\label{eq:rho}
\end{equation}
where $e_B^{\rm pred}$ is the unbiased ($z=0$) forecast made for that block. Later forecasts at $d$ are multiplied by $\rho_d^B$. Dimensions that have not yet been used keep $\rho=1$.

\begin{algorithm}[t]
\caption{Risk-aware feed-forward dimension control}\label{alg:ff}
\begin{algorithmic}[1]
\State Block 0: probe at $d_0=4$ (no key); $\sigma\gets\sigma_0$; $\rho_d^{Z,X}\gets1$ for all $d$
\For{$j=1,2,\dots$}
\State From public data of block $j-1$: $\hat x\gets$ \textsc{Invert}$(d_{j-1},\hat e_Z,N,b)$ \Comment{\eqref{eq:inv1}--\eqref{eq:inv2}}
\State Update $\sigma$ by \eqref{eq:sigma}; update $\rho_{d_{j-1}}$ by \eqref{eq:rho}
\For{$d'\in\mathcal D$, $z\in\{0,0.5,\dots,3\}$}
\State Forecast $(N,e_Z,e_X)$ at $(d',\hat x e^{z\sigma})$ \Comment{\eqref{eq:fc}}
\State $(e_Z,e_X)\gets(\rho_{d'}^Z e_Z,\rho_{d'}^X e_X)$; thresholds by \eqref{eq:th}
\State $\ell(d',z)\gets$ \eqref{eq:ell}; $V(d',z)\gets\ell\,\Phi(z)$
\EndFor
\State Announce $\theta_j=\arg\max V$; run block $j$ as fixed-length $\Pi_{\theta_j}$
\EndFor
\end{algorithmic}
\end{algorithm}

\begin{table}[t]
\caption{Controller settings (not optimized)}\label{tab:hyper}
\centering\footnotesize\setlength{\tabcolsep}{4pt}
\begin{tabular}{@{}ll@{}}
\toprule
Probe dimension $d_0$ & 4 \\
Count derating & $0.97$ \\
Sampling margin / absolute margin & $3$ s.d. / 0.002 \\
Pessimism grid $z$ & $\{0,0.5,\dots,3\}$ \\
Initial / floor of $\sigma$ & 0.15 / 0.03 \\
EWMA weight for $\sigma$ & 0.2 \\
Recalibration weight / clip range & 0.3 / $[0.5,3]$ \\
Inversion grid & $D/r_0\in[0,6]$, 2401 points \\
\bottomrule
\end{tabular}
\end{table}

\subsection{Baselines}\label{sec:baselines}
Five strategies are compared on the same channel realizations. The fixed-$d$ strategies keep the dimension constant but use the same risk-aware planner as the controller, so the comparison reflects only the choice of dimension. The naive planner assumes that the next block will be identical to the previous one ($z=0$, $\sigma=0$) and uses only statistical margins. The risk-aware controller is Algorithm~\ref{alg:ff} without recalibration ($\rho\equiv1$), and the risk-aware controller with recalibration is the complete algorithm. Finally, the oracle is a non-causal reference that knows the true expected count and error rates of every block, chooses the best $d$, and plans with the same margins. No strategy obtains key from block 0.

\section{Link and Channel Model}\label{sec:model}
\subsection{Turbulence-Induced Crosstalk}
We model turbulence as a thin Kolmogorov phase screen $\phi$ at the receiver aperture~\cite{Paterson2005}. The modes are
\begin{equation}
\mathrm{LG}_0^{\ell}(\rho,\varphi)=\sqrt{\frac{2}{\pi|\ell|!}}\frac{1}{w_0}\Big(\frac{\sqrt2\rho}{w_0}\Big)^{|\ell|}e^{-\rho^2/w_0^2}e^{i\ell\varphi},
\end{equation}
with waist $w_0=0.32R$ for aperture radius $R=D/2$. Screens are generated by the fast-Fourier-transform method from the phase power spectral density
\begin{equation}
\Phi_\phi(f)=0.023\,r_0^{-5/3}f^{-11/3},
\end{equation}
on a $512^2$ grid cropped to $256^2$, with eight levels of subharmonics to restore low-order statistics~\cite{Lane1992,Schmidt2010}. On a validation set of 300 screens, the simulated phase structure function is 0.94--0.99 times the Kolmogorov value $D_\phi(r)=6.88(r/r_0)^{5/3}$ for separations of 2 to 64 pixels (up to about one quarter of the aperture diameter), with the largest deficit at the largest separation. We compensate by scaling the phase by 1.04, after which the structure function lies within 7\% of theory over the same range.

For each screen we compute the overlap matrix over the 16-mode set,
\begin{equation}
M_{\ell'\ell}=\int_{\rm ap}\mathrm{LG}_{\ell'}^*\,e^{i\phi}\,\mathrm{LG}_{\ell}\,dA,\qquad \ell,\ell'\in S_{16},
\end{equation}
and obtain $M^{(d)}$ for smaller $d$ as the sub-block on $S_d$. For an input basis state $b_j$ and same-basis measurement $B$ (columns $b_k$), the outcome probabilities are
\begin{equation}
P(k\mid j)=\big|(B^\dagger M^{(d)}B)_{kj}\big|^2,\label{eq:pkj}
\end{equation}
and the remainder $1-\sum_kP(k\mid j)$ is lost from the mode set. The channel error rate $e_B^{\rm ch}$ is the pooled off-diagonal fraction, and $\eta$ is the pooled in-set transmission. The fast-steering mirror is modeled by removing the least-squares piston, tip, and tilt over the aperture, which accounts for most of the Kolmogorov phase variance~\cite{Noll1976}. We pool 400 screens per turbulence level $x=D/r_0\in[0,6]$ and interpolate linearly between levels.

\subsection{Atmosphere and Pass Geometry}
The aircraft flies at altitude $H=3$\,km and speed $v=120$\,m\,s$^{-1}$ along a track with horizontal offset $Y_0=2$\,km from the ground station. The pass lasts 160\,s and is centered on the time of closest approach, $t_c=80$\,s, so that the slant range and zenith angle $\zeta$ satisfy
\begin{equation}
L(t)=\sqrt{v^2(t-t_c)^2+Y_0^2+H^2},\qquad \sec\zeta=L/H.
\end{equation}
The refractive-index structure parameter follows the Hufnagel--Valley profile~\cite{AndrewsPhillips2005}
\begin{multline}
C_n^2(h)=0.00594\Big(\frac{v_w}{27}\Big)^2(10^{-5}h)^{10}e^{-h/1000}\\+2.7\times10^{-16}e^{-h/1500}+A\,e^{-h/100},
\end{multline}
with $v_w=21$\,m\,s$^{-1}$ and ground term $A$. The Fried parameter is the plane-wave expression
\begin{equation}
r_0=\Big[0.423\,k^2\sec\zeta\,\Gamma(t)\int_0^H C_n^2(h)\,dh\Big]^{-3/5},
\end{equation}
at $\lambda=2\pi/k=1550$\,nm, with receiver aperture $D=10$\,cm. The factor $\Gamma(t)=\exp(\sigma_{\ln}z_t-\sigma_{\ln}^2/2)$ models log-normal fluctuations of the integrated turbulence, where $z_t$ is a unit-variance first-order autoregressive process with coefficient $e^{-T/\tau_c}$, $\sigma_{\ln}=0.5$, and correlation time $\tau_c=15$\,s.

\subsection{Count Statistics}
The expected signal coincidences in a block are $S=R_0\,g(t)\,\eta(d,x)\,T$, where $R_0=2\times10^6$\,s$^{-1}$ is the coincidence rate at the 3\,km reference range before mode filtering and $g(t)=\min[1,(3\,{\rm km}/L)^2]$ accounts for geometric loss. Background coincidences arrive at $b$ counts\,s$^{-1}$ per mode, adding $N_b=d\,b\,T$ counts per block that are uniformly distributed over outcomes. The observed error rate in basis $B$ is therefore
\begin{equation}
e_B=\frac{S\,e_B^{\rm ch}(d,x)+N_b(d-1)/d}{S+N_b}.\label{eq:mix}
\end{equation}
Turbulence is averaged over the ensemble within each block, since its coherence time (typically a few milliseconds) is much shorter than $T$. Total counts are drawn from a Poisson distribution and error counts from a binomial distribution. The scenarios and parameters are listed in Table~\ref{tab:params}. In the two model-mismatch scenarios, the true channel is generated without tip/tilt correction, while the controller's model assumes that tip/tilt correction is applied.

\begin{table}[t]
\caption{Simulation parameters}\label{tab:params}
\centering\footnotesize\setlength{\tabcolsep}{3pt}
\begin{tabular}{@{}ll@{}}
\toprule
Wavelength / Rx aperture / waist & 1550\,nm / 10\,cm / $w_0=0.32R$ \\
Aircraft altitude / offset / speed & 3\,km / 2\,km / 120\,m\,s$^{-1}$ \\
Pass duration / block $T$ & 160\,s / 1\,s \\
HV ground term $A$ (night, day) & $1.7\times10^{-14}$, $1.7\times10^{-13}$\,m$^{-2/3}$ \\
Background $b$ (night, day) & 30, 3000 counts\,s$^{-1}$\,mode$^{-1}$ \\
Dawn & log-linear ramp night$\to$day \\
Turbulence fluctuation & $\sigma_{\ln}=0.5$, $\tau_c=15$\,s \\
Coincidence rate at 3\,km range & $2\times10^{6}$\,s$^{-1}$ (before mode filter) \\
$p_Z$, $f$, $\epsec$, $\epcor$ & 0.9, 1.16, $10^{-10}$, $10^{-15}$ \\
Menu $\mathcal D$ & $\{2,4,8,16\}$ \\
Monte-Carlo passes per scenario & 50 \\
\bottomrule
\end{tabular}
\end{table}

\section{Hardware-in-the-Loop Emulation}\label{sec:hw}
\subsection{Encoding the Link on a Processor}
The simulation includes only the imperfections that are explicitly modeled. To test the protocol against additional, unmodeled errors, we reproduce the measurement statistics of the link on a superconducting quantum processor. Hilder \emph{et al.}~\cite{Hilder2025} emulated qubit BB84 and BBM92, including attacks and noise, on a trapped-ion processor. Here we emulate qudit links with turbulence and use the measured device errors in the controller study.

For a screen $M$, input $b_j$, and basis $B$, let $w=B^\dagger M^{(d)}b_j$, with $\|w\|\le1$ by passivity. We prepare the $(\log_2 d+1)$-qubit state
\begin{equation}
|\psi\rangle=\sum_{k=0}^{d-1}w_k|k\rangle|0\rangle_{\rm f}+\sqrt{1-\|w\|^2}\,|0\rangle|1\rangle_{\rm f}\label{eq:psi}
\end{equation}
and measure all qubits in the computational basis. Outcomes with flag $1$ are counted as loss. By construction, the ideal outcome statistics equal those of the link in \eqref{eq:pkj}, including loss from the mode set. The processor adds its own state-preparation, gate, and readout errors, which play the role of an imperfect state analyzer.

\subsection{Experiment}
States were compiled with the \texttt{State\-Preparation} routine of Qiskit~\cite{Shende2006,Qiskit2024} at optimization level 3. On \texttt{ibm\_kingston} (IBM Heron), the mean two-qubit gate counts after transpilation were \twoqTwo, \twoqFour, and \twoqEight\ for $d=2,4,8$. We ran $D/r_0\in\{0,1,2,3\}$ with tip/tilt correction, using 12 screens per turbulent level and one screen at $D/r_0=0$, both bases, all inputs, and $d\in\{2,4,8\}$. That is $37\times2\times(2+4+8)=1036$ circuits at 256 shots each, submitted as a single SamplerV2 job (job \texttt{\hwJob}; 1.22 IBM runtime units).

\subsection{Device Offsets}
For each $(d,B)$ and turbulence level, we compare the measured error rate $e^{\rm hw}$ with the ideal error rate $e^{\rm id}$ computed from \eqref{eq:psi} on the \emph{same} screens, which removes screen-sampling noise from the comparison. Confidence intervals are 95\% Wilson score intervals~\cite{Wilson1927}. The device offset is the mean excess over turbulence levels,
\begin{equation}
\delta_{d,B}=\frac{1}{4}\sum_{x\in\{0,1,2,3\}}\big[e^{\rm hw}_{d,B}(x)-e^{\rm id}_{d,B}(x)\big].\label{eq:offset}
\end{equation}
In the hardware-in-the-loop passes of Section~\ref{sec:hil}, $\delta_{d,B}$ is added to $e_B^{\rm ch}(d,x)$ in both the true channel and the controller's model. The emulation reproduces measurement statistics only; photon transport, detector dead time, and multipair emission are not modeled.

\section{Results}\label{sec:results}
\subsection{Channel Error Rates and Design Maps}
\begin{figure}[t]
\centering\includegraphics[width=\columnwidth]{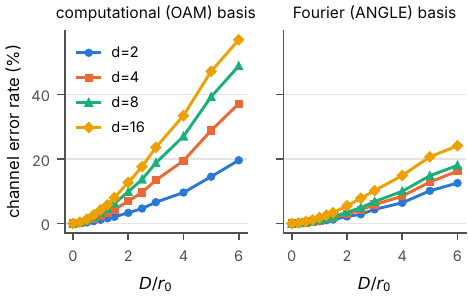}
\caption{Channel error rates in the key (OAM) and test (Fourier) bases versus turbulence strength $D/r_0$, with tip/tilt correction (400 screens per point).}\label{fig:channel}
\end{figure}

The pooled channel error rates are plotted in Fig.~\ref{fig:channel}. Errors increase with both $d$ and $D/r_0$, and they are considerably lower in the Fourier basis than in the OAM basis. With tip/tilt removed, turbulence mostly scatters light into neighboring OAM orders. This scattering resembles a shift in OAM, and since the Fourier-basis states are eigenstates of the cyclic shift, it produces relatively few errors in that basis. The same ordering was found in independent simulations~\cite{Jaouni2024}.

\begin{figure*}[t]
\centering\includegraphics[width=0.9\textwidth]{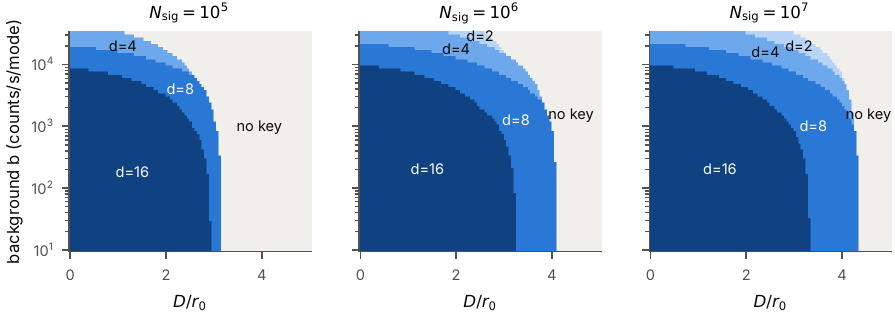}
\caption{Finite-key-optimal dimension for a known channel versus turbulence $D/r_0$ and background $b$, for $N_{\rm sig}$ signal coincidences per block. The signal rate is $10^6$\,s$^{-1}$, so $N_{\rm sig}$ also sets the integration time for background counts. Light gray: no positive key.}\label{fig:map}
\end{figure*}

Fig.~\ref{fig:map} shows the dimension that maximizes \eqref{eq:ell} when the channel is known and constant. Under night-like conditions ($b\lesssim10^2$\,s$^{-1}$, $D/r_0\lesssim2$), and well beyond them, $d=16$ is optimal. With stronger turbulence or more background light, the optimum decreases to $d=8$ and then to $d=4$; $d=2$ is optimal only in a narrow band at high background and appears only for $N_{\rm sig}\ge10^6$. Beyond these regions no key can be extracted. Larger blocks extend the region with positive key because the sampling correction $\mu$ in \eqref{eq:mu} decreases. No single dimension is optimal over the range of conditions encountered in a pass, and the boundaries between regions also depend on the block size. A fixed-dimension system has to commit to one point in this space before deployment.

\subsection{Airborne Passes}
Table~\ref{tab:pass} lists the total secret key over \NSEED\ Monte-Carlo passes of \nBlocks\ blocks each, and Table~\ref{tab:ops} gives the corresponding operating statistics. At night and at dawn, $d=16$ is the best fixed dimension, and the feed-forward controller obtains \nightAdaptFrac\% and \dawnAdaptFrac\% of its key. At night the controller switches about once per pass, from the probe dimension to $d=16$. In daylight it switches about \daySwitches\ times per pass, mostly between $d=4$ and $d=8$, and produces \dayGainPct\% more key than the best fixed dimension. Over identical channel realizations, the mean per-pass ratio is $1.096\pm0.011$ (standard error), and the controller produced at least as much key as the best fixed dimension in all 50 daytime passes. The controller does not need to know in advance which fixed dimension is best. Fixing $d=4$ would have cost \dayFourLoss\% of the daytime key, and fixing $d=8$ would have cost \nightEightLoss\% of the night key.

\begin{table*}[t]
\caption{Secret key per pass (Mbit), mean over \NSEED\ passes; feed-forward (FF) is the risk-aware controller without recalibration (cf.\ Table~\ref{tab:variants}). Bold: best fixed dimension. Ratios are of the means.}\label{tab:pass}
\centering\footnotesize
\begin{tabular}{@{}lcccccccc@{}}
\toprule
Scenario & $d{=}2$ & $d{=}4$ & $d{=}8$ & $d{=}16$ & Feed-forward ($\pm$s.d.) & Oracle & FF / best fixed & FF / oracle \\
\midrule
Night & 40.6 & 98.9 & 159.1 & \textbf{216.4} & 215.9$\pm$7.1 & 233.1 & 0.998 & 0.93 \\
Dawn & 29.9 & 74.2 & 118.8 & \textbf{155.6} & 155.8$\pm$13.7 & 178.1 & 1.001 & 0.88 \\
Day & 12.4 & 31.3 & 46.6 & \textbf{47.2} & 50.8$\pm$16.7 & 68.8 & 1.075 & 0.74 \\
Night, model mismatch & 22.6 & 59.1 & \textbf{83.7} & 80.6 & 80.5$\pm$18.7 & 127.9 & 0.962 & 0.63 \\
Dawn, model mismatch & 7.9 & 23.0 & \textbf{29.0} & 18.5 & 19.9$\pm$12.0 & 52.2 & 0.686 & 0.38 \\
\bottomrule

\end{tabular}
\end{table*}

\begin{table}[t]
\caption{Operating statistics: fraction of the 160 blocks that yield key (probe included) and dimension switches per pass, mean of \NSEED\ passes}\label{tab:ops}
\centering\footnotesize\setlength{\tabcolsep}{3.5pt}
\begin{tabular}{@{}lcccc@{}}
\toprule
 & \multicolumn{3}{c}{Blocks with key} & \\
\cmidrule(lr){2-4}
Scenario & Best fixed & FF & Oracle & Switches \\
\midrule
Night & 0.975 & 0.973 & 0.994 & 1.0 \\
Dawn & 0.803 & 0.837 & 0.898 & 4.5 \\
Day & 0.384 & 0.521 & 0.659 & 12.7 \\
Night, mismatch & 0.832 & 0.688 & 0.954 & 24.9 \\
Dawn, mismatch & 0.433 & 0.320 & 0.628 & 33.0 \\
\bottomrule
\end{tabular}
\end{table}

\begin{figure}[!t]
\centering\includegraphics[width=\columnwidth]{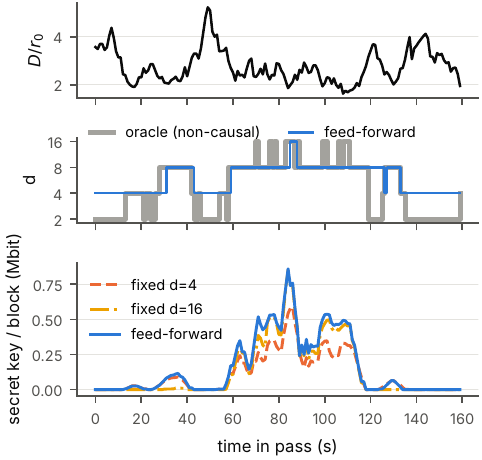}
\caption{One daytime pass. Top: turbulence $D/r_0$. Middle: dimension chosen by the causal feed-forward controller and by the non-causal oracle. Bottom: secret key per block (5-block moving average) for feed-forward and two fixed dimensions.}\label{fig:pass}
\end{figure}

The daytime gain can be understood from Table~\ref{tab:ops}. The best fixed dimension, $d=16$, produces key in only 38\% of the blocks because it fails during the parts of the pass with strong turbulence and long range. The controller produces key in 52\% of the blocks by switching to $d=4$ or $d=8$ during those parts and returning to larger $d$ when conditions improve. An example daytime pass is shown in Fig.~\ref{fig:pass}. Since it relies on a forecast, the controller tends to follow the oracle with some delay at transitions and to be more conservative when the turbulence increases quickly.

The oracle provides an upper bound on what a controller with perfect knowledge of the channel could achieve. The feed-forward controller reaches 93\% of the oracle's key at night, 88\% at dawn, and \dayOracleFrac\% in daylight. Section~\ref{sec:kalman} examines whether a better forecaster would reduce this difference.

\subsection{Block Duration}
Table~\ref{tab:block} shows the effect of the block duration $T$ in daylight. Short blocks incur larger finite-size penalties, whereas long blocks average over channel variations that occur within the block and are harder to forecast. For causal operation, the total key is largest at $T=1$\,s among the durations tested; the oracle, which does not forecast, continues to improve as $T$ increases. The controller outperforms the best fixed dimension for all three block durations; the paired ratios decrease slightly with $T$, but the differences are within about two standard errors.

\begin{table}[t]
\caption{Daytime key per pass (Mbit) versus block duration. Paired: mean per-pass ratio FF/best fixed $\pm$ standard error.}\label{tab:block}
\centering\footnotesize\setlength{\tabcolsep}{3pt}
\begin{tabular}{@{}lccccc@{}}
\toprule
$T$ (s) & Best fixed & FF & Oracle & Paired & Switches \\
\midrule
0.25 & 37.1 ($d{=}16$) & 39.8 & 47.4 & 1.123$\pm$0.024 & 16.1 \\
1 & 47.2 ($d{=}16$) & 50.8 & 68.8 & 1.096$\pm$0.011 & 12.7 \\
4 & 31.8 ($d{=}8$) & 34.6 & 80.6 & 1.072$\pm$0.015 & 6.3 \\
\bottomrule

\end{tabular}
\end{table}

\subsection{Planner Ablation and Robustness}\label{sec:ablation}
\begin{figure*}[t]
\centering\includegraphics[width=0.92\textwidth]{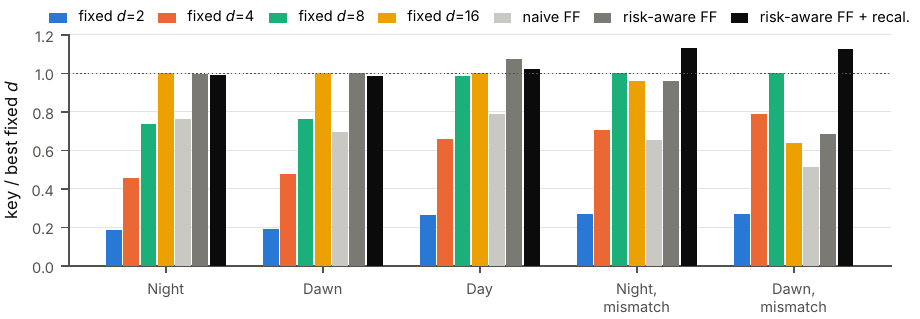}
\caption{Key per pass relative to the best fixed dimension of each scenario (dotted line), mean of \NSEED\ passes. Every fixed dimension falls to 73.5\% or less in at least one scenario; the recalibrated feed-forward controller never falls below \recalWorst\%.}\label{fig:robust}
\end{figure*}

Table~\ref{tab:variants} and Fig.~\ref{fig:robust} show the contribution of each component of the controller. The naive planner loses a large share of its blocks to aborts, and the risk-aware planner produces \raGainMin\%--\raGainMax\% more key in every scenario. Of the components tested, accounting for forecast uncertainty has the largest effect.

In the mismatch scenarios, the channel is generated without tip/tilt correction while the controller assumes it is present. This does not affect security (Proposition~\ref{prop:ff}), but it does affect performance. Without recalibration, the controller overestimates the benefit of large $d$, repeatedly chooses dimensions whose error rates it underestimates, aborts, and switches again; Table~\ref{tab:ops} shows 25--33 switches per pass and fewer successful blocks than the best fixed dimension. At dawn it obtains only 69\% of the key of the best fixed dimension. With recalibration, the controller learns the higher error rates after a few blocks at each dimension and produces about \recalMMgainMax\% more key than the best fixed dimension. When the model is correct, recalibration costs \recalCostMin\%--\recalCostMax\% of the key, because its correction factors are estimated from noisy data and sometimes overestimate the error rates.

\begin{table}[t]
\caption{Controller variants: key per pass (Mbit), mean of \NSEED\ passes. Bottom row: worst ratio to the best fixed dimension over all five scenarios.}\label{tab:variants}
\centering\footnotesize\setlength{\tabcolsep}{3pt}
\begin{tabular}{@{}lccccc@{}}
\toprule
Scenario & Naive & Risk-aware & + recal. & Best fixed & Oracle \\
\midrule
Night & 165.1 & 215.9 & 215.1 & 216.4 ($d{=}16$) & 233.1 \\
Dawn & 108.3 & 155.8 & 153.7 & 155.6 ($d{=}16$) & 178.1 \\
Day & 37.4 & 50.8 & 48.4 & 47.2 ($d{=}16$) & 68.8 \\
Night, mismatch & 54.8 & 80.5 & 94.6 & 83.7 ($d{=}8$) & 127.9 \\
Dawn, mismatch & 15.0 & 19.9 & 32.7 & 29.0 ($d{=}8$) & 52.2 \\
\midrule
Worst case / best fixed & 0.52 & 0.69 & \textbf{0.988} & 1 & -- \\
\bottomrule

\end{tabular}
\end{table}

The last row of Table~\ref{tab:variants} summarizes these results. In all five scenarios, the recalibrated controller obtains at least \recalWorst\% of the key of the best fixed dimension for that scenario. In contrast, each fixed dimension falls to \fixBestWorst\% or less in at least one scenario: $d=16$ drops to \fixSixteenWorst\% under model mismatch, and $d=8$ to 73.5\% at night. Since the best fixed dimension is not known before deployment, this worst-case performance is the main practical advantage of adaptation.

\subsection{Limits of Forecasting}\label{sec:kalman}
The difference between the controller and the oracle may be due to the forecaster or to turbulence that cannot be predicted from past data. To distinguish these, we replaced the EWMA forecaster with a two-state Kalman filter that is given the flight geometry and the true turbulence statistics, which is the most favorable case for the filter. The state is $s=(u,m)$, where $u=\ln x-\tfrac35\ln\sec\zeta$ is the log-turbulence corrected for geometry and $m$ is its slowly varying mean. The variable $u$ follows the same first-order autoregressive dynamics as the simulated atmosphere, with coefficient $e^{-T/\tau_c}$ and process noise variance $(0.6\sigma_{\ln})^2(1-e^{-2T/\tau_c})$, and $m$ follows a random walk with a standard deviation of 0.02 per block. The measurement variance in each block is obtained from the binomial uncertainty of $\hat e_Z$ by applying the delta method to \eqref{eq:inv2}. The one-step prediction of the filter and its standard deviation replace $\hat x$ and $\sigma$ in Algorithm~\ref{alg:ff}.

\begin{table}[t]
\caption{Kalman-filter versus EWMA forecaster (\NSEED\ passes each; key in Mbit). RMSE: one-block-ahead forecast error in $\ln(D/r_0)$.}\label{tab:kalman}
\centering\footnotesize\setlength{\tabcolsep}{2.5pt}
\begin{tabular}{@{}lccccc@{}}
\toprule
Scenario & EWMA & Kalman & Ratio ($\pm$s.e.) & Kalman wins & RMSE (E / K) \\
\midrule
Night & 215.9 & 218.0 & 1.010$\pm$0.002 & 80\% & 0.106 / 0.105 \\
Dawn & 155.8 & 156.7 & 1.005$\pm$0.002 & 64\% & 0.107 / 0.107 \\
Day & 50.8 & 51.1 & 1.007$\pm$0.005 & 60\% & 0.106 / 0.105 \\
\bottomrule
\end{tabular}
\end{table}

As Table~\ref{tab:kalman} shows, the Kalman filter increases the total key by only \kfGainMin\%--\kfGainMax\%, and the root-mean-square error of the one-block-ahead forecast of $\ln(D/r_0)$ is about 0.106 for both forecasters. The remaining gap is thus caused by turbulence fluctuations that are unpredictable from one block to the next, rather than by the estimator. For $\tau_c=15$\,s and $T=1$\,s, the innovation standard deviation of $\ln C_n^2$ is $\sigma_{\ln}\sqrt{1-e^{-2T/\tau_c}}\approx0.18$, which corresponds to about 0.11 in $\ln x$ and agrees with the observed forecast error. Reducing the gap would require the key length to depend on the data of the same block, as in a variable-length protocol~\cite{Tupkary2024,KanitscharPRApplied2025}, and not a better forecaster.

\subsection{Processor Emulation}
\begin{figure}[t]
\centering\includegraphics[width=\columnwidth]{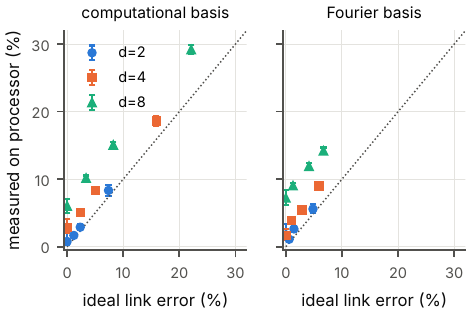}
\caption{Processor-measured versus ideal link error (\texttt{ibm\_kingston}; 95\% Wilson intervals; four turbulence levels per dimension). Dotted: equality.}\label{fig:hw}
\end{figure}

Table~\ref{tab:hw} and Fig.~\ref{fig:hw} compare the error rates measured on the processor with the ideal error rates for the same screens. For every dimension, the measured error increases with turbulence and follows the ideal values. Linear fits of the measured against the ideal error rates have slopes between \hwSlopeMin\ and \hwSlopeMax, and the excess error varies by no more than \hwSpread\ percentage points across turbulence levels. The device errors can therefore be described as an approximately constant additive offset, $\delta_{d,Z}/\delta_{d,X}$, of \offTwoZ/\offTwoX\% at $d=2$, \offFourZ/\offFourX\% at $d=4$, and \offEightZ/\offEightX\% at $d=8$. The increase with $d$ is consistent with the larger circuits (on average \twoqTwo, \twoqFour, and \twoqEight\ two-qubit gates) and the larger number of measured qubits at higher $d$. The fraction of shots kept decreases with turbulence in agreement with the in-set transmission, which confirms that the flag qubit correctly reproduces the loss of the link.

\begin{table}[t]
\caption{Error rates (\%) measured on \texttt{ibm\_kingston}. In parentheses: ideal link error for the same screens. Last row: fraction of shots kept (flag $0$).}\label{tab:hw}
\centering\footnotesize\setlength{\tabcolsep}{4pt}
\begin{tabular}{@{}cccccc@{}}
\toprule
 & & \multicolumn{4}{c}{$D/r_0$} \\
\cmidrule(l){3-6}
$d$ & Basis & 0 & 1 & 2 & 3 \\
\midrule
2 & Z & 0.8 (0.0) & 1.7 (1.2) & 2.9 (2.3) & 8.3 (7.4) \\
2 & X & 1.8 (0.0) & 1.1 (0.6) & 2.6 (1.4) & 5.6 (4.8) \\
4 & Z & 2.8 (0.0) & 5.0 (2.4) & 8.3 (5.1) & 18.6 (15.9) \\
4 & X & 1.7 (0.0) & 3.8 (1.0) & 5.4 (2.9) & 9.0 (5.9) \\
8 & Z & 6.0 (0.0) & 10.1 (3.4) & 15.0 (8.2) & 29.2 (22.1) \\
8 & X & 7.2 (0.0) & 9.0 (1.3) & 11.9 (4.1) & 14.2 (6.7) \\
\midrule
\multicolumn{2}{@{}l}{Kept} & 0.99 & 0.94 & 0.86 & 0.75 \\
\bottomrule

\end{tabular}
\end{table}

\subsection{Hardware in the Loop}\label{sec:hil}
To assess the effect of these device errors on the protocol, we added the offsets to the channel model, provided the same offsets to the controller, and repeated the passes with the dimensions run on the processor, $\mathcal D=\{2,4,8\}$ (Table~\ref{tab:loop}). The device errors reduce the daytime key by \loopDayDrop\%. They also change which fixed dimension is best: in daylight the optimum changes from $d=8$ to $d=4$, since the $d=8$ offset uses up much of the additional error tolerance of $d=8$. At dawn, $d=8$ remains the best fixed dimension, but its advantage over $d=4$ falls from 60\% to 9\%. The controller adapts to these changes without any retuning and, in every scenario, comes within a few tenths of a percent of the best fixed dimension or exceeds it. A dimension chosen for ideal devices would therefore have been the wrong choice for daytime operation with these devices.

\begin{table}[t]
\caption{Hardware-in-the-loop passes, $\mathcal D=\{2,4,8\}$ (Mbit). ``Model + processor'' adds the offsets measured in Table~\ref{tab:hw}.}\label{tab:loop}
\centering\footnotesize\setlength{\tabcolsep}{3pt}
\begin{tabular}{@{}llcccccc@{}}
\toprule
 & Channel & $d{=}2$ & $d{=}4$ & $d{=}8$ & FF & Oracle & FF/best \\
\midrule
Night & model only & 40.6 & 98.9 & 159.1 & 159.0 & 171.2 & 1.000 \\
Night & model + processor & 33.3 & 71.9 & 83.2 & 83.4 & 90.9 & 1.002 \\
Dawn & model only & 29.9 & 74.2 & 118.8 & 119.0 & 134.7 & 1.001 \\
Dawn & model + processor & 24.1 & 52.2 & 57.0 & 58.0 & 67.4 & 1.018 \\
Day & model only & 12.4 & 31.3 & 46.6 & 46.8 & 61.8 & 1.003 \\
Day & model + processor & 9.3 & 18.9 & 14.0 & 19.0 & 25.7 & 1.002 \\
\bottomrule

\end{tabular}
\end{table}

\section{Discussion}\label{sec:discussion}
\subsection{Separation of Security and Performance}
With feed-forward selection, no additional security analysis is required. Each block is a standard fixed-length protocol, so security follows from composition (Proposition~\ref{prop:ff}), and the overall security parameter increases only linearly with the number of blocks (Corollary~\ref{cor:pass}). As a result, the controller may be heuristic, learned from data, or based on a physical model, and it can be modified later without repeating the security analysis. The same reasoning applies to other parameters, such as the basis bias, the block length, or decoy intensities in prepare-and-measure protocols, as long as they are set from earlier public data.

\subsection{Gap to the Oracle}
Because the controller must forecast the channel, it cannot match the oracle. Section~\ref{sec:kalman} shows that this gap is mainly due to unpredictable turbulence and not to the choice of estimator. The gap could be reduced in two ways. Shorter blocks follow the channel more closely but incur larger finite-size penalties (Table~\ref{tab:block}). Alternatively, a variable-length protocol could set the key length after the errors of the block are known, while the dimension is still chosen in advance~\cite{Tupkary2024}. The dimension determines which measurements are made and must be set before the block; the key length does not have this restriction.

\subsection{Benefit of Adaptation}
The main benefit of adaptation is robustness. The best fixed dimension is different at night and in daylight, with ideal and with imperfect devices, and with and without tip/tilt correction. The recalibrated controller follows these changes without prior information and in the worst case retains \recalWorst\% of the key of the best fixed dimension (Table~\ref{tab:variants}). For a link that must operate under conditions that cannot all be characterized in advance, this worst-case behavior matters more than the average gain in any one scenario.

\subsection{Implementation Considerations}
Because the mode sets $S_d$ are nested, the $Z$-basis measurement at any $d\le16$ uses the same sorter outputs, and changing $d$ only changes which outcomes are kept. The $X$-basis measurement requires projections onto superpositions within $S_d$. Programmable devices such as digital micromirror devices, which have been used to switch between OAM-MUB states at kilohertz rates~\cite{Mirhosseini2015}, can be reconfigured well within the time between blocks. For $T=1$\,s, the time needed to announce $\theta_j$ (a few bytes) and to run the controller (a search over 4 dimensions and 7 risk levels) is negligible. As with all public communication in QKD, the announcements must be authenticated. The turbulence estimate is computed from the public error rates, so no additional optical measurements are needed. The controller does, however, need an estimate of the background rate $b$, which in practice can be obtained from coincidences in time windows adjacent to the coincidence window; in our simulations the controller is given the true value of $b$.

\subsection{Limitations}\label{sec:limits}
This study has several limitations. (i) The finite-key bound \eqref{eq:ell} assumes exact MUB measurements, basis-independent detection, and trusted, characterized devices, and error correction is represented by a constant efficiency $f$. Multipair emission, detector dead time, and side channels are not included. (ii) Turbulence is modeled as a thin phase screen at the receiver, without scintillation or beam wander, and the channel is averaged within each block. (iii) The pass geometry and background levels are representative values, not measurements, and the controller is given the true background rate rather than an estimate. (iv) The processor emulation reproduces measurement statistics but not photon transport; the device errors stand in for imperfections of the state analyzer and were measured only for $d\le8$. (v) The controller settings (Table~\ref{tab:hyper}) were not optimized, and recalibration was tested against only one type of model mismatch. (vi) The security argument relies on existing qudit uncertainty-relation bounds~\cite{Tomamichel2012,Bradler2016}, and the key-length formula and Proposition~\ref{prop:ff} should be checked independently by QKD security specialists before deployment. The method is not limited to OAM encoding. Time-bin and time--frequency encodings, for which the optimal dimension also depends on operating conditions~\cite{Chang2026,BergmayrMann2026}, could use the same controller with an appropriate channel model.

\section{Conclusion}\label{sec:conclusion}
We have shown that the dimension of an HD-QKD protocol can be changed during operation without a new security proof. When the dimension and key length of each block are set from public data of earlier blocks, existing fixed-length proofs compose, and errors in the channel model affect only the key rate. In simulations of an airborne pass, accounting for forecast uncertainty increased the key by \raGainMin\%--\raGainMax\% compared with a naive planner. With online recalibration, the controller obtained at least \recalWorst\% of the key of the best fixed dimension in every scenario, including channel-model mismatch, without knowing in advance which dimension that was. A comparison with a Kalman-filter forecaster indicated that the remaining gap to the oracle is due to unpredictable turbulence and not to the estimator. Measurements on a superconducting processor showed that realistic device errors change the optimal dimension and that the controller adapts to this change. Future work will address a variable-length extension to reduce the forecasting gap and an optical free-space demonstration.

\appendices
\section{Reproducibility}\label{app:repro}
All simulations use fixed pseudorandom seeds. For pass $s\in\{0,\dots,49\}$, the turbulence time series is generated from seed $s$ and is shared by all strategies, so that comparisons are paired; the block-level Poisson and binomial draws of each strategy use separate seeded streams. The channel table (400 screens at each turbulence level, with and without tip/tilt correction) is generated once and reused. The processor experiment used the circuit inputs and job identifier given in Section~\ref{sec:hw}. The analysis pipeline consists of (1) building the channel table, (2) running the pass simulations and ablations, (3) submitting and analyzing the processor job, and (4) generating all reported numbers and tables directly from the saved results.

\section*{Acknowledgment}
The authors acknowledge the use of IBM Quantum services for this work. The views expressed are those of the authors and do not reflect the official policy or position of IBM or the IBM Quantum team. The authors used an AI assistant (Claude, Anthropic) during code development and manuscript preparation and take full responsibility for the content.

\section*{Conflict of Interest}
The authors declare no conflicts of interest.

\section*{Code and Data Availability}
The simulation and analysis code, the processor circuit inputs, and the measured counts are available from the corresponding author upon reasonable request.

\bibliographystyle{IEEEtran}
\bibliography{refs}

@inproceedings{BB84,
  author = {C. H. Bennett and G. Brassard},
  title = {Quantum cryptography: Public key distribution and coin tossing},
  booktitle = {Proc. IEEE Int. Conf. Computers, Systems and Signal Processing},
  address = {Bangalore, India}, pages = {175--179}, year = {1984}}

@article{Cerf2002,
  author = {N. J. Cerf and M. Bourennane and A. Karlsson and N. Gisin},
  title = {Security of quantum key distribution using $d$-level systems},
  journal = {Phys. Rev. Lett.}, volume = {88}, pages = {127902}, year = {2002}}

@article{Sheridan2010,
  author = {L. Sheridan and V. Scarani},
  title = {Security proof for quantum key distribution using qudit systems},
  journal = {Phys. Rev. A}, volume = {82}, pages = {030301(R)}, year = {2010}}

@article{TomamichelRenner2011,
  author = {M. Tomamichel and R. Renner},
  title = {Uncertainty relation for smooth entropies},
  journal = {Phys. Rev. Lett.}, volume = {106}, pages = {110506}, year = {2011}}

@article{Tomamichel2012,
  author = {M. Tomamichel and C. C. W. Lim and N. Gisin and R. Renner},
  title = {Tight finite-key analysis for quantum cryptography},
  journal = {Nat. Commun.}, volume = {3}, pages = {634}, year = {2012}}

@article{Islam2017,
  author = {N. T. Islam and C. C. W. Lim and C. Cahall and J. Kim and D. J. Gauthier},
  title = {Provably secure and high-rate quantum key distribution with time-bin qudits},
  journal = {Sci. Adv.}, volume = {3}, pages = {e1701491}, year = {2017}}

@article{PortmannRenner2022,
  author = {C. Portmann and R. Renner},
  title = {Security in quantum cryptography},
  journal = {Rev. Mod. Phys.}, volume = {94}, pages = {025008}, year = {2022}}

@article{Cozzolino2019,
  author = {D. Cozzolino and B. {Da Lio} and D. Bacco and L. K. Oxenl{\o}we},
  title = {High-dimensional quantum communication: Benefits, progress, and future challenges},
  journal = {Adv. Quantum Technol.}, volume = {2}, pages = {1900038}, year = {2019}}

@article{Erhard2020,
  author = {M. Erhard and M. Krenn and A. Zeilinger},
  title = {Advances in high-dimensional quantum entanglement},
  journal = {Nat. Rev. Phys.}, volume = {2}, pages = {365--381}, year = {2020}}

@article{KanitscharPRL2025,
  author = {F. Kanitschar and M. Huber},
  title = {Practical framework for analyzing high-dimensional quantum key distribution setups},
  journal = {Phys. Rev. Lett.}, volume = {135}, pages = {010802}, year = {2025}}

@article{KanitscharPRApplied2025,
  author = {F. Kanitschar and M. Huber},
  title = {Composable finite-size security of high-dimensional quantum-key-distribution protocols},
  journal = {Phys. Rev. Appl.}, volume = {24}, number = {5}, pages = {054028}, year = {2025}}

@phdthesis{KanitscharThesis2025,
  author = {F. Kanitschar},
  title = {Quantum Communication Beyond Qubits},
  school = {TU Wien}, address = {Vienna, Austria}, year = {2025}}

@article{Chang2026,
  author = {K.-C. Chang and M. C. Sarihan and N. K. H. Li and F. Kanitschar and K. E. Akyuz and Y. Chen and D.-I. Lee and J. H. Kang and A. Aldhafeeri and A. Mueller and M. D. Shaw and B. Korzh and M. Spiropulu and P. Erker and M. Huber and C. W. Wong},
  title = {Unlocking the potential of high-dimensional quantum communication with scalable photonic entanglement in time and frequency},
  journal = {Sci. Adv.}, volume = {12}, number = {27}, pages = {eaee1333}, year = {2026}}

@article{Bulla2023,
  author = {L. Bulla and M. Pivoluska and K. Hjorth and O. Kohout and J. Lang and S. Ecker and S. P. Neumann and J. Bittermann and R. Kindler and M. Huber and M. Bohmann and R. Ursin},
  title = {Nonlocal temporal interferometry for highly resilient free-space quantum communication},
  journal = {Phys. Rev. X}, volume = {13}, pages = {021001}, year = {2023}}

@article{Ecker2021,
  author = {S. Ecker and B. Liu and J. Handsteiner and M. Fink and D. Rauch and F. Steinlechner and T. Scheidl and A. Zeilinger and R. Ursin},
  title = {Strategies for achieving high key rates in satellite-based {QKD}},
  journal = {npj Quantum Inf.}, volume = {7}, pages = {5}, year = {2021}}

@inproceedings{Ecker2023space,
  author = {S. Ecker and J. Pseiner and J. Piris and M. Bohmann},
  title = {Advances in entanglement-based {QKD} for space applications},
  booktitle = {Proc. SPIE 12777, Int. Conf. Space Optics (ICSO 2022)}, pages = {1277727}, year = {2023}}

@article{Tupkary2024,
  author = {D. Tupkary and E. Y.-Z. Tan and N. L{\"u}tkenhaus},
  title = {Security proof for variable-length quantum key distribution},
  journal = {Phys. Rev. Res.}, volume = {6}, pages = {023002}, year = {2024}}

@article{Paterson2005,
  author = {C. Paterson},
  title = {Atmospheric turbulence and orbital angular momentum of single photons for optical communication},
  journal = {Phys. Rev. Lett.}, volume = {94}, pages = {153901}, year = {2005}}

@book{Schmidt2010,
  author = {J. D. Schmidt},
  title = {Numerical Simulation of Optical Wave Propagation with Examples in {MATLAB}},
  publisher = {SPIE Press}, address = {Bellingham, WA}, year = {2010}}

@article{Lane1992,
  author = {R. G. Lane and A. Glindemann and J. C. Dainty},
  title = {Simulation of a {K}olmogorov phase screen},
  journal = {Waves in Random Media}, volume = {2}, number = {3}, pages = {209--224}, year = {1992}}

@book{AndrewsPhillips2005,
  author = {L. C. Andrews and R. L. Phillips},
  title = {Laser Beam Propagation through Random Media},
  edition = {2nd}, publisher = {SPIE Press}, address = {Bellingham, WA}, year = {2005}}

@article{Mafu2013,
  author = {M. Mafu and A. Dudley and S. Goyal and D. Giovannini and M. McLaren and M. J. Padgett and T. Konrad and F. Petruccione and N. L{\"u}tkenhaus and A. Forbes},
  title = {Higher-dimensional orbital-angular-momentum-based quantum key distribution with mutually unbiased bases},
  journal = {Phys. Rev. A}, volume = {88}, pages = {032305}, year = {2013}}

@article{Mirhosseini2015,
  author = {M. Mirhosseini and O. S. Maga{\~n}a-Loaiza and M. N. O'Sullivan and B. Rodenburg and M. Malik and M. P. J. Lavery and M. J. Padgett and D. J. Gauthier and R. W. Boyd},
  title = {High-dimensional quantum cryptography with twisted light},
  journal = {New J. Phys.}, volume = {17}, pages = {033033}, year = {2015}}

@misc{Dziwulska2026,
  author = {K. Dziwulska and C. Spiess and S. Mishra and M. Leipe and Y. Hadiyal and F. Steinlechner},
  title = {Self-stabilized high-dimensional quantum key distribution on a metropolitan free-space link},
  howpublished = {arXiv:2602.22102}, year = {2026}}

@misc{BergmayrMann2026,
  author = {A. E. Bergmayr-Mann and G. Murta and M. Huber},
  title = {Temporal trade-offs in high-dimensional entanglement: a comprehensive noise model for optimal time-bin {QKD} protocols},
  howpublished = {arXiv:2609.35488}, year = {2026}}

@article{Scarfe2025,
  author = {L. Scarfe and F. Hufnagel and M. F. Ferrer-Garcia and A. D'Errico and K. Heshami and E. Karimi},
  title = {Fast adaptive optics for high-dimensional quantum communications in turbulent channels},
  journal = {Commun. Phys.}, volume = {8}, pages = {79}, year = {2025}}

@inproceedings{Muller2023,
  author = {R. M{\"u}ller and D. Bacco and L. K. Oxenl{\o}we and S. Forchhammer},
  title = {Information reconciliation for high-dimensional quantum key distribution using nonbinary {LDPC} codes},
  booktitle = {Proc. Int. Symp. Topics in Coding (ISTC)}, year = {2023}, note = {doi: 10.1109/ISTC57237.2023.10273570}}

@article{Shende2006,
  author = {V. V. Shende and S. S. Bullock and I. L. Markov},
  title = {Synthesis of quantum-logic circuits},
  journal = {IEEE Trans. Comput.-Aided Des. Integr. Circuits Syst.}, volume = {25}, number = {6}, pages = {1000--1010}, year = {2006}}

@misc{Qiskit2024,
  author = {A. Javadi-Abhari and others},
  title = {Quantum computing with {Q}iskit},
  howpublished = {arXiv:2405.08810}, year = {2024}}

@article{Bradler2016,
  author = {K. Br{\'a}dler and M. Mirhosseini and R. Fickler and A. Broadbent and R. Boyd},
  title = {Finite-key security analysis for multilevel quantum key distribution},
  journal = {New J. Phys.}, volume = {18}, pages = {073030}, year = {2016}}

@article{WangLo2019,
  author = {W. Wang and H.-K. Lo},
  title = {Machine learning for optimal parameter prediction in quantum key distribution},
  journal = {Phys. Rev. A}, volume = {100}, pages = {062334}, year = {2019}}

@misc{OptiQKD2026,
  author = {N. Mohamed and J. Kaldari and S. Al-Kuwari},
  title = {{AML-QKD}: Adaptive machine learning framework for real-time parameter tuning in {QKD}},
  howpublished = {arXiv:2603.04192}, year = {2026}}

@misc{Bae2023,
  author = {M. J. Bae and N. K. Panigrahy and P. Dhara and W. O. Krawec and A. Russell and D. Towsley and B. Wang},
  title = {Blockwise key distillation in satellite-based quantum key distribution},
  howpublished = {arXiv:2307.04300}, year = {2023}}

@article{Sidhu2022,
  author = {J. S. Sidhu and T. Brougham and D. McArthur and R. G. Pousa and D. K. L. Oi},
  title = {Finite key effects in satellite quantum key distribution},
  journal = {npj Quantum Inf.}, volume = {8}, pages = {18}, year = {2022}}

@misc{ArqandTan2025,
  author = {A. Arqand and E. Y.-Z. Tan},
  title = {Marginal-constrained entropy accumulation theorem},
  howpublished = {arXiv:2502.02563}, year = {2025}}

@phdthesis{TupkaryThesis2026,
  author = {D. Tupkary},
  title = {Rigorous Security Proofs for Practical Quantum Key Distribution},
  school = {University of Waterloo}, address = {Waterloo, Canada}, year = {2026}, note = {arXiv:2604.21791}}

@misc{Hilder2025,
  author = {J. Hilder and S. Heu{\ss}en and A. Ginter and A. Wilke and L. Postler and U. Poschinger and F. Schmidt-Kaler and W. Wormsbecher},
  title = {Quantum internet in a nutshell -- advancing quantum communication with ion traps},
  howpublished = {arXiv:2507.14383}, year = {2025}}

@article{Jaouni2024,
  author = {T. Jaouni and L. Scarfe and F. Bouchard and M. Krenn and K. Heshami and F. {Di Colandrea} and E. Karimi},
  title = {Predicting atmospheric turbulence for secure quantum communications in free space},
  journal = {Opt. Express}, volume = {33}, number = {5}, pages = {10759}, year = {2025}}

@misc{Bhat2025,
  author = {V. S. Bhat and R. Chatterjee and K. Bajar and S. Mujumdar},
  title = {Optimizing the qudit dimensions of position-momentum entangled photons for {QKD}},
  howpublished = {arXiv:2505.00811}, year = {2025}}

@article{BBM92,
  author = {C. H. Bennett and G. Brassard and N. D. Mermin},
  title = {Quantum cryptography without {B}ell's theorem},
  journal = {Phys. Rev. Lett.}, volume = {68}, pages = {557--559}, year = {1992}}

@article{Scarani2009,
  author = {V. Scarani and H. Bechmann-Pasquinucci and N. J. Cerf and M. Du{\v{s}}ek and N. L{\"u}tkenhaus and M. Peev},
  title = {The security of practical quantum key distribution},
  journal = {Rev. Mod. Phys.}, volume = {81}, pages = {1301--1350}, year = {2009}}

@article{Pirandola2020,
  author = {S. Pirandola and others},
  title = {Advances in quantum cryptography},
  journal = {Adv. Opt. Photon.}, volume = {12}, number = {4}, pages = {1012--1236}, year = {2020}}

@phdthesis{Renner2005,
  author = {R. Renner},
  title = {Security of Quantum Key Distribution},
  school = {ETH Zurich}, address = {Zurich, Switzerland}, year = {2005}, note = {arXiv:quant-ph/0512258}}

@article{Allen1992,
  author = {L. Allen and M. W. Beijersbergen and R. J. C. Spreeuw and J. P. Woerdman},
  title = {Orbital angular momentum of light and the transformation of {L}aguerre--{G}aussian laser modes},
  journal = {Phys. Rev. A}, volume = {45}, pages = {8185--8189}, year = {1992}}

@article{Berkhout2010,
  author = {G. C. G. Berkhout and M. P. J. Lavery and J. Courtial and M. W. Beijersbergen and M. J. Padgett},
  title = {Efficient sorting of orbital angular momentum states of light},
  journal = {Phys. Rev. Lett.}, volume = {105}, pages = {153601}, year = {2010}}

@article{Krenn2014,
  author = {M. Krenn and R. Fickler and M. Fink and J. Handsteiner and M. Malik and T. Scheidl and R. Ursin and A. Zeilinger},
  title = {Communication with spatially modulated light through turbulent air across {V}ienna},
  journal = {New J. Phys.}, volume = {16}, pages = {113028}, year = {2014}}

@article{Nauerth2013,
  author = {S. Nauerth and F. Moll and M. Rau and C. Fuchs and J. Horwath and S. Frick and H. Weinfurter},
  title = {Air-to-ground quantum communication},
  journal = {Nat. Photonics}, volume = {7}, pages = {382--386}, year = {2013}}

@article{Pugh2017,
  author = {C. J. Pugh and S. Kaiser and J.-P. Bourgoin and J. Jin and N. Sultana and S. Agne and E. Anisimova and V. Makarov and E. Choi and B. L. Higgins and T. Jennewein},
  title = {Airborne demonstration of a quantum key distribution receiver payload},
  journal = {Quantum Sci. Technol.}, volume = {2}, pages = {024009}, year = {2017}}

@article{Liu2021drone,
  author = {H.-Y. Liu and others},
  title = {Optical-relayed entanglement distribution using drones as mobile nodes},
  journal = {Phys. Rev. Lett.}, volume = {126}, pages = {020503}, year = {2021}}

@article{Liao2017,
  author = {S.-K. Liao and others},
  title = {Satellite-to-ground quantum key distribution},
  journal = {Nature}, volume = {549}, pages = {43--47}, year = {2017}}

@article{Yin2017,
  author = {J. Yin and others},
  title = {Satellite-based entanglement distribution over 1200 kilometers},
  journal = {Science}, volume = {356}, pages = {1140--1144}, year = {2017}}

@article{Noll1976,
  author = {R. J. Noll},
  title = {Zernike polynomials and atmospheric turbulence},
  journal = {J. Opt. Soc. Am.}, volume = {66}, number = {3}, pages = {207--211}, year = {1976}}

@article{Wilson1927,
  author = {E. B. Wilson},
  title = {Probable inference, the law of succession, and statistical inference},
  journal = {J. Am. Stat. Assoc.}, volume = {22}, number = {158}, pages = {209--212}, year = {1927}}

\begin{IEEEbiographynophoto}{Murtuza Lemonwala} received the B.S. degree in aerospace engineering, with a double major in mechanical engineering, from Embry-Riddle Aeronautical University, Prescott, AZ, USA, and the M.S. degree in mechanical engineering from Embry-Riddle Aeronautical University, Daytona Beach, FL, USA. He is currently pursuing the Ph.D. degree in electrical engineering and computer science with the Department of Electrical Engineering and Computer Science, Embry-Riddle Aeronautical University, Daytona Beach. His research interests include applied quantum computing and artificial intelligence for aerospace and autonomous systems, and quantum communication over free-space links.
\end{IEEEbiographynophoto}

\begin{IEEEbiographynophoto}{Laxima Niure Kandel} received the M.S. degree in telecommunication and information systems from the University of Essex, Colchester, U.K., and the Ph.D. degree in computer engineering from Stevens Institute of Technology, Hoboken, NJ, USA, in 2020, where she received the Outstanding Dissertation Award for her doctoral work.

She is currently an Assistant Professor with the Department of Electrical Engineering and Computer Science, Embry-Riddle Aeronautical University, Daytona Beach, FL, USA, where she directs the Cybersecurity and Assured System Engineering (CyBASE) Laboratory. Her research interests include anomaly and fault detection using machine learning and artificial intelligence, physical-layer security of unmanned aerial vehicles, GPS spoofing mitigation, information security, and WiFi localization and sensing.
\end{IEEEbiographynophoto}

\EOD
\end{document}